\documentclass[twocolumn,aps,secnumarabic,balancelastpage,amsmath,amssymb,nofootinbib]{revtex4}

\usepackage{listings}
\usepackage{chapterbib}
\usepackage{color}
\usepackage{graphics}
\usepackage[pdftex]{graphicx}
\usepackage{longtable}
\usepackage{epsf}
\usepackage{bm}
\usepackage{thumbpdf}
\usepackage[colorlinks=true]{hyperref}
\usepackage{algorithm}
\usepackage{algorithmic}

\newcommand{\e}{\varepsilon}

\newcommand{\sd}{Schr\"{o}dinger }

\begin{document}
\title{Robustness of Controlled Quantum Dynamics}
\author{Andy Koswara}
\affiliation{School of Chemical Engineering, Purdue University}
\author{Raj Chakrabarti}
\thanks{Corresponding author}
\email{raj@pmc-group.com}
\affiliation{School of Chemical Engineering, Purdue University\\ Center for Advanced Process Decision Making, Department of Chemical Engineering, Carnegie Mellon University}
\altaffiliation{Current address: Division of Fundamental Research, PMC Advanced Technology LLC, Mt. Laurel, NJ 08054}
\date{\today}

\begin{abstract}
Control of multi-level quantum systems is sensitive to implementation errors in the control field and uncertainties associated with system Hamiltonian parameters. A small variation in the control field spectrum or the system Hamiltonian can cause an otherwise optimal field to deviate from controlling desired quantum state transitions and reaching a particular objective. An accurate analysis of robustness is thus essential in understanding and achieving model-based quantum control, such as in control of chemical reactions based on \emph{ab initio} or experimental estimates of the molecular Hamiltonian. In this paper, theoretical foundations for quantum control robustness analysis are presented from both a distributional perspective - in terms of moments of the transition amplitude, interferences, and transition probability - and a worst-case perspective. Based on this theory, analytical expressions and a computationally efficient method for determining the robustness of coherently controlled quantum dynamics are derived. The robustness analysis reveals that
there generally exists a set of control pathways that are more resistant to destructive interferences in the presence of control field
and system parameter uncertainty. These robust pathways interfere and combine to yield a relatively accurate transition amplitude and high transition probability when uncertainty is present.
\end{abstract}
\maketitle
%
\section{Introduction}
\label{Section1}
The study of quantum dynamics controlled by an external field has rapidly developed over the past 30 years \cite{leth1977, bloe1984, warr1993, levi2001, brif2010,brif2010b}. Here, the \textit{quantum control} objective is typically to design an external electromagnetic field which coherently manipulates a quantum system from an initial state to a desired final state. The appropriate engineering of a control field has been implemented in various quantum settings leading to several important  technological applications, such as selective bond dissociation of compounds \cite{brix2001}, discrimination of similar biomolecules \cite{peter2010}, real-time microscopy of biological systems \cite{min2012}, and quantum computing \cite{ahar1999}. 

Many of the aforementioned examples of successful control of quantum systems have been achieved using model-free experimental learning control techniques \cite{brif2010b}. While model-based control of low-dimensional systems, such as nuclear spin states, has been performed successfully, limitations in field generation and shaping technology and imperfect knowledge of the system render model-based control of higher-dimensional systems (for e.g. molecular ro-vibrational states) more challenging \cite{mabu2005}. Model-based control of quantum dynamics has been studied in the presence of various types of uncertainty in both the system Hamiltonian and the manipulated control field \cite{pryo2006,li2006,prav2003,kosu2013,grac2012,belt2011,hock2014}.
In particular, it is infeasible to perfectly model the Hamiltonian of a large quantum system since \textit{ab initio} methods become computationally intractable without some approximation and because laboratory measurements are not readily accessible. An example is the case of determining the vibrational energies of a polytatomic molecule, whereby a particular bond is analyzed in isolation and the rest of the molecule is treated as a disturbance with bounded energy \cite{beum1992}.
Similarly, a time-varying control field can either be subject to uncertainty due to stochastic fluctuations in either time or frequency domain field variables, or due to inaccuracies in the values of manipulated field parameters.  In the context of laser control, for example, these can originate due to perturbations of laser sources in the laboratory and the limited precision of laser pulse-shaping technology \cite{wein2000, jian2007}.  When designing the profile of a control field for optimizing a quantum performance criterion, such factors must be taken into account in order to ensure quantum control \textit{robustness}. Control of quantum dynamics that maintains high fidelity in the presence of these types of uncertainty is referred to as robust quantum control \cite{wese2004,wang2012,grac2012,kosu2013, cabr2011}.

Various approaches to quantification of robustness have been proposed in the engineering literature
\cite{ma1999, nagy2004, fan1991, ferr1997}, with the majority being based on leading order Taylor expansions of the control performance measure. Robustness of control is generally expressed in terms of moments of the distribution of the performance measure $J$ \cite{nagy2004}, or the distance between the nominal performance measure and its worst-case value ($J_{wc}$) \cite{ma1999}. There are several methods for approximating the latter in the presence of uncertainty \cite{ma1999, fan1991, ferr1997}. These are often based on solving constrained optimization problems using computationally efficient algorithms \cite{braa1996}. Early studies on quantum control robustness
described the robustness of controlled dynamics qualitatively in terms of the effect of control and system uncertainty distributions on the dynamical trajectory \cite{toth1994, rabi2002}.  For example, one study described how phase noise reduces the control pulse area and, in turn, the population transfer as shifts in the spectral frequency lead to inefficient resonance and lack of constructive pathway interferences \cite{toth1994}. Another study \cite{rabi2002} examined the inherent degree of robustness in an optimal control field due to the bilinearity of quantum observable expectation values in the evolution operator and its adjoint. More recent studies on quantum robust control have introduced several types of numerical approximations such as leading order expansions in order to quantify robustness \cite{belt2011,grac2012,kosu2013,heul2010,hock2014}.

In engineering control, leading order Taylor expansions are commonly applied in conjunction with real-time feedback control that corrects for deviations between the desired and actual trajectories.
In the absence of real-time feedback (which is currently impossible for many important quantum systems),
leading order Taylor expansions can be inaccurate in the prediction of the moments of state variables. Moreover, such leading order approximations do not provide a mechanistic understanding of how robustness can be achieved in terms of the underlying dynamical pathways responsible for control fidelity.

In this work, we present an asymptotic approach to quantification of quantum control robustness that is accurate with respect to calculation of the first and second moments (and higher moments if desired) of the performance measure.
We provide a general asymptotic theory for computation and control of moments of bilinear quantum systems in the presence of Hamiltonian and control field uncertainty,  without relying on linearization or related leading order Taylor expansions.
The robustness of the quantum dynamics is analyzed in terms of implementation errors
in the classical input variables (in a semiclassical picture of controlled quantum dynamics) and parameter uncertainty in the quantum Hamiltonian.
The method calculates the effect of uncertainty in the control field and in the system's dipole moment on the fidelity of control. In addition, different quantum pathways involved in the controlled dynamics are delineated such that qualitative and quantitative analysis of robustness may be more precisely discussed in terms of the moments of interferences between different order pathways and their contributions to the transition amplitude and probability.

The paper is organized as follows: Section \ref{Section2} describes the theory of quantum control via combination and interference of quantum pathways. In Section \ref{Section3}, methods for characterization of uncertainty in the system Hamiltonian and control field are briefly presented as a starting point in quantum control robustness analysis. The procedure for calculating the robustness criteria is then presented in Section \ref{Section4}. Here, an example of how the robustness analysis is carried out assuming Gaussian uncertainty distributions is described. In Section \ref{Section5}, the numerical implementation of the robustness analysis method is described and its application on control of an four-level Hamiltonian is demonstrated in Section \ref{Section6}. Here, the potential use of the robustness analysis theory in the development of robust control algorithms and in aiding laboratory learning control is also discussed. We finally conclude with a summary and future work in Section \ref{Section7}.

\section{Quantum Control via Multiple Pathway Interference}
\label{Section2}
\hspace{0.2in}
In a semiclassical picture of a controlled quantum system coupled with a time-varying external field, the dynamics can be described by the \sd equation:
\begin{gather}
	\frac{dU(t)}{dt}=-\frac{i}{\hbar}\left(H_0-\mu\varepsilon(t)\right)U(t),~ U(0)=I,
	\label{SDunit}
\end{gather}
where $H_0$ is the time-independent Hamiltonian of the system, $\mu$ the dipole moment, $\varepsilon(t)$ the time dependent field, and $U(t)$ denotes the unitary propagator. In order to allow for a simplified notation in the ensuing analysis, the notation for the interaction Hamiltonian $H_I(t) = e^{\frac{i}{\hbar}H_0t}\{-\mu\e(t)\}e^{-\frac{i}{\hbar}H_0t}$ is used, giving:
\begin{gather}
	\frac{dU_{I}(t)}{dt}=-\frac{i}{\hbar}H_{I}(t)U_{I}(t),~ U_I(t)=e^{\frac{i}{\hbar}H_0t}U(t).
	\label{SDunitinter}
\end{gather}	
In general, the quantum control objectives can be categorized into two types: (i) population transfer control (i.e.  $|U_{ji}(T)|^2\rightarrow 1$) such as in chemical reaction control, or (ii) dynamical propagator control (i.e. $U(T)\rightarrow U_{target}$) for use in quantum computation. The work described herein applies to robustness analysis of both control categories. It is important to note that since $|U_{ji}(T)|^2=|U_{I,ji}(T)|^2$, and that $U_I(t)$ can be readily inverse-transformed to $U(t)$ according to (\ref{SDunitinter}) the subscript $I$ is dropped from the description of the unitary propagator in the interaction picture for convenience.

The transition amplitude $U_{ji}(T)$ can be calculated as a sum of an infinite Dyson series \cite{dyso1951}:
\begin{align}
	U_{ji}&(T)= \langle j |\bigg[ \sum_{m=1}^{\infty} \left(-\frac{\imath}{\hbar}\right)\int_0^T H_I(t_{1})~dt_1+ \nonumber \\
	& \left(-\frac{\imath}{\hbar}\right)^2 \int_0^T \int_0^{t_2} H_I(t_{1})~H_I(t_{2})~dt_1~dt_2+ \cdots \nonumber \\
	& +\left(-\frac{\imath}{\hbar}\right)^m\int_0^T\int_0^{t_2} \cdots \int_0^{t_{m-1}}  H_I(t_{1})~H_I(t_{2}) \cdots \nonumber \\
	&~~~~~H_I(t_m) ~dt_1~dt_2\cdots dt_{m}\bigg]| i \rangle.
	\label{DysSer}
\end{align}
We use the notation $U_{ji}^m(T)$ to denote the $m$-th order term in the series above.  Quantum interferences occur due to coherence terms $(U_{ji}^mU_{ji}^{m'*})$ in the expression for the transition probability $P_{ji}=|U_{ji}(T)|^2$.  \emph{Constructive} interference corresponds to $\mathrm{Re}\left\{(U_{ji}^{m}(T))(U_{ji}^{m'}(T))^*\right\}$
is larger than 0, and \emph{destructive} interference corresponds to values less than 0.

A fundamental concept in the theory of quantum control robustness analysis that we will develop and apply in this work is a \emph{quantum pathway}. Prior work has considered the characterization of quantum pathways in the context of quantum control mechanism analysis \cite{mitr2003,mitr2008}. In the context of robustness analysis, a quantum pathway is a term in the Dyson series expansion written in terms of the products of the form $\prod^K_{k=1} x_k^{\alpha_k}$, where $x$ (or its log) denotes either a control variable or a time-independent Hamiltonian parameter.  This includes the conventional multiphoton pathways (or combinations thereof) as well as other types of pathways as will be described below. Like multiphoton pathways, the other types of quantum pathways can interfere to produce the observed dynamics.

Using a cosine representation of a control field with $K$ spectral modes, $\e(t)=\sum_{k}^K A(\omega_k)\left(\mathrm{cos}(\omega_{k}t)+\phi(\omega_k)\right)$ for an N-dimensional quantum system the transition amplitude $U_{ji}(T)$ can be expressed as a function of the field's spectral parameters and the system's dipole operator elements:
\begin{widetext}
\begin{align}
	U_{ji}&(T) = \sum_{m} \left(\frac{\imath}{\hbar}\right)^m \sum_{k_{m}=1}^K A_{k_m} \sum_{l_{m-1}=1}^N \mu_{jl_{m-1}} \int_0^T e^{\imath\left(\omega_{jl_{m-1}}t_{m}\right)} \cos(\omega_{k_m}t_{m} + \phi(\omega_{k_m})) \times \cdots \nonumber \\
	&\times \sum_{k_{1}=1}^K A_{k_1}  \sum_{l_1=1}^N \mu_{l_1i} \int_0^{t_{2}}e^{\imath\left(\omega_{l_1i}t_{1}\right)} \cos(\omega_{k_1}t_{1} + \phi(\omega_{k_1})) \nonumber \\
	&~~~~~~~~~~~~~~~~~~~dt_1\cdots dt_{m}.
	\label{dysonserparam}
\end{align}
\end{widetext}
In the above, the shorthand notations $A(\omega_k)=A_k$ and $\omega_{ji}=\frac{(E_j-E_i)}{\hbar}$ have been used. The control and system parameters in (\ref{dysonserparam}) may be sorted in a way that the transition amplitude can be interpreted as a sum of \textit{quantum pathways}. For example, the  transition amplitude may be rewritten as:
\begin{align}
	&U_{ji}(T) = \sum_{m} \left(\frac{\imath}{\hbar}\right)^m \sum_{\vec \alpha \in \mathcal{M}} \prod_{k=1}^K A_k^{\alpha_k} \times \nonumber\\
	&\sum_{(k_1,\cdots,k_m)} \sum_{l_{m-1}}^N \mu_{jl_{m-1}} \int_0^T e^{i\omega_{jl_{m-1}}t_m}\cos(\omega_{k_m}t_m + \phi(\omega_{k_m})) \times \nonumber\\
	&\cdots\times \sum_{l_{1}}^N \mu_{l_11}\int_0^{t_2} e^{i\omega_{l_1i}t_1}\cos(\omega_{k_1}t_1 + \phi(\omega_{k_1})) ~ dt_1 \cdots dt_m, \label{amppath}
\end{align}
where the sum $\sum_{(k_1,\cdots,k_m)}$ is over all $1 \leq k_i \leq K,~i=1,\cdots,m$ such that mode $k$ appears in the multiple integral $\alpha_k$ times.
In (\ref{amppath}) the ${m-th}$ order Dyson term is expressed as a sum of terms with powers of amplitude $[\alpha_1,\cdots,\alpha_K]$ such that $\sum_{k=1}^K {\alpha_k=m}$. The notation $\vec{\alpha} \in \mathcal{M}$ is used to denote all such pathways belonging to a particular order $m$ i.e. the integer polytope  $\mathcal{M} \equiv \{\vec{\alpha} \in \mathbb{Z}^K~|~\sum_{k=1}^K {\alpha_k=m,~\vec \alpha > 0}\}$. In this way, the transition amplitude can be described as a sum of \emph{amplitude pathways}, in which each pathway is denoted by a unique combination of $\vec{\alpha}=[\alpha_1,\cdots,\alpha_K]$:
\begin{align}
	U_{ji}(T)=\sum_{m} U_{ji}^m(T) = \sum_{m} \sum_{\vec{\alpha}\in \mathcal{M}} U_{ji}\left(T,\vec{\alpha}\right).
	\label{transamppath}
\end{align}
For example, given two modes (i.e. ${K=2}$), the two $1$st order amplitude pathways can be identified as $U_{ji}^1(T,\vec{\alpha}=[0,1])$ and $U_{ji}^1(T,\vec{\alpha}=[1,0])$. Analogously, the $m$-th order transition amplitude may be described as a sum of \emph{dipole pathways}, which are more commonly known as multiphoton transition pathways. The dipole pathways are given as follows:
\begin{align}
	&U_{ji}(T,\vec \alpha) = \left(\frac{\imath}{\hbar}\right)^m ~~\prod_{p<q} \mu_{pq}^{\alpha_{pq}} \times
\nonumber \\
	&\sum_{(l_1,\cdots,l_{m-1})} \sum_{k_m}^K A_{k_m} \int_0^T e^{i\omega_{jl_{m-1}}t_m}\cos(\omega_{k_m}t_m + \phi(\omega_{k_m})) \times \nonumber\\
	&\cdots \times \sum_{k_1}^K A_{k_1}\int_0^{t_2} e^{i\omega_{l_1i}t_1}\cos(\omega_{k_1}t_1 + \phi(\omega_{k_1}))~dt_1 \cdots dt_m,
\label{dippath}
\end{align}
where $\vec \alpha \in  \mathcal{M}$, and the sum $\sum_{(l_1,\cdots,l_{m-1})}$ is over all $1 \leq l_i \leq N,~i=1,\cdots,m-1$ such that frequency $\pm \omega_{pq}$ corresponding to dipole parameters $\mu_{pq},\mu_{qp}$ appears in the multiple integral $\alpha_{pq}$ times. Phase pathways will be considered in a separate work.

Given the definition of quantum pathways as associated with the control and system parameters, the quantum control robustness can be precisely described. Indeed, the robustness may be defined in terms of how distribution and magnitude of variations in the field's amplitude and phase parameters and system's dipole moments change the trajectory of the quantum pathways and, hence, the transition amplitude and probability. Extending from the concepts of \cite{ferr1997, nagy2004, ma1999, fan1991, braa1996}, we can express as robustness criteria the moments of the (quantum pathway) interferences and transition amplitude and probability. This will be described in Section \ref{Section4}. In the following Section \ref{Section3}, the statistical description of the uncertainties associated with the system's Hamiltonian parameters and control field
are presented.
%

\section{Characterization of System Parameter and Control Field Uncertainty}
\label{Section3}
The robustness criteria introduced in the previous Section \ref{Section2} can be computed once system parameter and control input uncertainties
have been characterized. In robust control engineering, one is concerned with the effects of uncertainty in the time-independent parameters characterizing the equations of motion of the system, as well as disturbances or implementation errors in input variables \cite{nagy2004,braa1996}. The former parameters are not directly observable, whereas the latter variables are generally observable.

Hamiltonian parameter estimation is achieved through system identification based on measurements of the observed dynamics. Hamiltonian parameter estimates can be obtained by either frequentist (e.g., maximum likelihood, ML) or Bayesian estimation techniques.
For illustration, we consider uncertainty in a dipole operator $\mu$ that is real and has diagonal elements equal to zero; this operator can be parameterized by the vector $\vec \theta = [\mu_{12},\cdots,\mu_{(N-1)N}]^T$ of independent elements $\mu_{pq},~q>p$ of which there are at most $\frac{N^2-N}{2}$. We assume that all elements $\mu_{pq},~q \neq p$ are uncertain and we denote by $K$ the number of parameters.

Denoting by $L(\theta|x)$ the 
likelihood function (a function of $\theta$) for dipole parameter estimation \cite{chak2012} based on a set of measurements $x$, the maximum likelihood estimator $\hat \theta_{ML}= \mathrm{arg ~max} ~L({\theta}|x)$ is an asymptotically efficient estimator with the corresponding covariance matrix of parameter estimates given by
\begin{align}
	\Sigma = \mathcal{I}^{-1}(\hat \theta). 
	\label{fisher}
\end{align}	
where $\mathcal{I}(\theta)$ denotes the Fisher information matrix
\begin{align}
	\mathcal{I}(\theta )=-\mathrm{E}\left[\frac{\partial ^{2}\ln \left(L(\theta |x)\right)}{\partial \theta \partial \theta^{\prime }}\right].
\end{align}
$\mathcal{I}^{-1}(\theta_0 )$ (where $\theta_0$ denotes the true parameter vector) is called the \emph{Cramer-Rao lower bound (CRB)} for consistent estimators. The ML estimator asymptotically achieves this lower bound on the covariance matrix.

Alternatively, \emph{Bayesian} Hamiltonian estimation may be used \cite{Gelman2004}. Bayesian Hamiltonian estimation can employ \emph{ab initio} calculations along with experimental data to construct system parameter estimates $\hat \theta$. Bayesian estimation is based on the notion of a prior plausibility distribution $p~(\theta \mid I)$ on the space $\Theta$ of parameters, which is updated to a posterior distribution $p~(\theta \mid x \land I)$ based on the measurements $x$, through the relation
\begin{align}
	p~(\theta \mid x \land I)~ d \theta &= \frac {L~(x \mid \theta) ~p~(\theta \mid I)~ d \theta}{\int_{\Theta} L~(x\mid \theta) ~p~(\theta \mid I) ~d \theta}.
	\label{genpost}
\end{align}
Here, $I$ denotes the prior information set and the likelihood $L$ is written as the conditional probability of the measurement outcomes given $\theta$ in order to derive the posterior distribution by application of Bayes' rule. Using this approach, we would have \emph{ab initio} estimates for parameters represented by $p(\theta \mid I)$, in addition to the parametric model and observation law which provide the likelihood function.  In the following robustness analysis, we assume a multivariate normal approximation to the posterior distribution of $\theta$ is available either from frequentist (e.g., ML) or Bayesian estimation.

In laser control of molecular dynamics, which is the application of primary interest in the current work, uncertainties in the control field can originate in two ways: a) inaccuracies in the values of manipulated field parameters \cite{stef2007}; b) stochastic disturbances or noise in the realizations of input variables.
The control input $\e(t)$ is manipulated in the frequency domain through the magnitudes of spectral amplitudes $A(\omega)$ and/or phases $\phi(\omega)$ of the laser field.
Irrespective of whether the random variables $\delta \e(t)$ originate due to stochastic fluctuations in the realizations of these variables or inaccuracies in manipulated parameters, the expressions for the moments of state variables are equivalent, as will be discussed below; hence the theory of robustness presented herein is applicable to both problems. In the examples considered herein, we are primarily concerned with errors in the manipulated spectral amplitudes or phases of the laser field.

For robustness analysis in the presence of field uncertainty, the frequency domain covariance function is used instead of the covariance matrix (\ref{fisher}) of parameter estimates.
For illustrative purposes, in the present work we consider examples with uncertain spectral amplitudes
and deterministic phases (i.e., $\delta \phi(\omega)=0$), and assume there is no correlation between the different spectral amplitude random variables:
\begin{align}
\mathrm{E}\left[\delta A(\omega)\delta A(\omega')\right]=0, ~\omega'\neq \pm \omega.
\label{freqcorr}
\end{align}
The theory is, however, also directly applicable to the case with correlated uncertainty in the frequency domain.

Studies of quantum control robustness to stochastic disturbances characterized by a time domain correlation function have also been reported, for example, in \cite{hock2014}, which considered stationary field noise processes.
Based on the theory of Fourier transforms, there exists a one-to-one mapping between frequency and time domain representations of field noise processes:
\begin{align}
\int_0^T \exp(-i\omega t) \delta \e(t)~dt &=  \delta \e(\omega) \\
&= \delta \left[A(\omega)\exp\left(i\phi(\omega)\right)\right].
\label{ft_phase_amp}
\end{align}
The correlation function,
\begin{align}
\frac{1}{\sigma_t\sigma_{t'}}\mathrm{E}\left[\delta \e(t)\delta \e(t')\right],
\end{align}
where $\sigma_t$ denotes the standard deviation of $\delta \e(t)$,
can be calculated given sampled amplitude/phase variations,
and the frequency domain correlation function may be calculated from sampled time domain variations, for any field noise process (stationary or nonstationary). In the context of robustness to control field disturbances, the theory and methodologies developed herein are most conveniently applied to disturbances wherein the frequency domain correlation function is a physically natural representation, which is the case for intensity and phase noise in laser control \cite{wein2000, jian2007,pasc2006,ralp1999}.

\section{Robustness analysis}
\label{Section4}
\subsection{Formulation of robustness criteria}
\label{Section41}
The eventual goal of robustness analysis is to understand how a control field achieves robust transition amplitude and probability when uncertainty is present in the control field or the system parameters. We consider robustness of the control performance measure (e.g., transition probability) to variations $\delta \theta$ in the parameters. Assuming the covariance matrix of parameter estimates is available as in (\ref{fisher}), the posterior distribution of $\delta \theta$ is modeled as a multivariate normal distribution, i.e., $\theta \sim \mathcal{N}(\hat \theta, \Sigma)$. Through choice of a confidence level $c$, we can specify the set of possible realizations of $\delta \theta$ corresponding to that confidence level as:
\begin{gather}
	\Theta=\{\delta \theta~|~\delta \theta^T \Sigma^{-1} \delta \theta~\leq~\chi_K^2(c)\}, \nonumber \\
	\delta \theta=\theta- \hat\theta,
	\label{distrib}
\end{gather}
where $\chi^2_K(c)$ denotes the inverse cumulative distribution function of the chi square distribution with $K$ degrees of freedom, $K$ denoting the number of noisy or uncertain parameters.
The distribution of $\delta \theta$ can be used to estimate the corresponding distribution of the control performance measure $J$. Let $J=P_{ji}$, the transition probability between states $i$ and $j$, and consider the case of dipole operator uncertainty as an example. With a 1st order Taylor expansion, the only distribution function that can be derived is a normal distribution
\begin{gather}
	\label{Jdistrib}
	J \sim \mathcal{N} (J (\hat \theta), \sigma_J^2),
\end{gather}
with variance
\begin{align}
	\sigma_J^2 \approx \mathrm{Tr}\left[\Sigma \nabla_\theta J (\nabla_\theta J)^T\right]
	\label{firsttaylor},
\end{align}
where
\begin{align}
	[\nabla_\theta J]_k&= -\frac{\imath}{\hbar} \mathrm{Tr}\bigg\{[|i\rangle \langle i|, U^\dagger(T) |j\rangle \langle j| U(T)] \times \nonumber \\
	&\int_0^T U^\dagger (t) X_k \varepsilon(t) U(t)~dt \bigg\},
	\label{gradJ}	
\end{align}
and $X_k$ is the Hermitian matrix obtained by setting ${\theta_k=1},~{\theta_l=0}, ~{l \neq k}$ in $\mu(\theta)$.

An analogous representation of the variance of the performance measure is possible in the case of input field uncertainty
in terms of either a frequency or time domain representation of the gradient of the performance measure with respect to the field variables \cite{brif2010} and the correlation function in the respective domain. As noted above, the expressions are equivalent for either implementation inaccuracies or field disturbances; only the correlation functions depend on the application, with implementation inaccuracies often displaying less correlation.

With higher order Taylor expansions \cite{belt2011}, one cannot derive a distribution function for $J$ analytically, although various approximate numerical methods have been proposed \cite{nagy2004}.
Worst-case robustness analysis can be formulated either in terms of maximization of the magnitude of the performance measure deviation $\delta J$ subject to the inequality constraints on $\delta \theta$ in (\ref{distrib}), or directly in terms of an approximation to the pdf of $J$. The former approach is considered further in Section 4.3. In the latter approach, using as an example (\ref{Jdistrib}) as an approximation to the pdf and specifying a confidence level $c$, an estimate of $J_{wc}$ can be expressed as 
\begin{gather}
	J_{wc}=J(\hat \theta) + \delta J_{wc} =  J(\hat \theta) - \sqrt{2} \sigma_{J} ~\mathrm{erf}^{-1}(c).
\label{Jwc}
\end{gather}

Robustness of nonlinear systems is commonly examined from the perspective of linearized control system dynamics.
However, for many important quantum systems, like femtosecond molecular dynamics, real time feedback control is currently impossible. In the absence of feedback, control system linearization (as well as associated leading order Taylor expansions) can be inaccurate as a method for prediction of the moments of state variables - and hence robustness of observable quantities to parameter uncertainty and disturbances – since the variance of the state variable deviations increases rapidly with evolution time and the linearized system is no longer an accurate approximation to the true nonlinear system.
Methods such as feedforward control are computationally less intensive and quantum feedforward controllers have been proposed based on linearized control systems \cite{chak2012}.

Most quantum robust control strategies typically apply leading order approximations to quantify the robustness of the control fidelity to system parameter uncertainty or field disturbances \cite{grac2012,belt2011,hock2014}.
For example, \cite{grac2012} considered robustness of pulses for quantum gate operations in the presence of Hamiltonian parameter uncertainty and input field disturbances using an approach based on second order perturbation theory.  \cite{belt2011} analyzed  the Hessian curvature of the quantum control landscape for population transfer at its extrema and its effect on robustness of optimal quantum control to field disturbances. This is a second order Taylor expansion approach to quantum control robustness analysis applied to nominally optimal controls in order to assess their robustness to field disturbances. The effects of landscape curvature on controlled gate robustness were also studied in \cite{hock2014}.
These approaches are analogous to leading order methods applied previously in the engineering literature \cite{nagy2004,braa1999}.

Here, we present an asymptotic approach that can provide accurate estimates of the 1st and 2nd moments (and higher moments if desired) of $J$ suitable for use in either distributional or worst-case robustness criteria for controlled quantum dynamics. This approach is more accurate than methods for moment calculations (like (\ref{firsttaylor})) based on leading order Taylor expansions.
In addition, following from the analysis of the \sd equation (\ref{dysonserparam}) and their interpretation as quantum pathways as in (\ref{amppath}) and (\ref{dippath}) one can determine how 
input and system parameter uncertainties explicitly affect the dynamical mechanism of controlled dynamics. Given an accurate description of the parameter distribution, its contribution of to each pathway and subsequently the transition amplitude and probability can be determined asymptotically up to a significant Dyson order $M$. Analogous to classical robust control \cite{ferr1997, nagy2004, ma1999, fan1991, braa1996}, given the noisy distribution of the input parameters, a measure of robustness can be expressed in terms of the moments of the quantum control objective (commonly, first and second).

However, unlike in the classical control counterpart, there is an interference phenomenon which is responsible for the observed dynamics in a quantum system, since the total transition probability between an initial state $|i\rangle$ and a final state $|j\rangle$ at time $T$ can be expressed as:
\begin{align}
	P_{ji}&=
	\sum_{m} \underbrace{\left|U_{ji}^m(T)\right|^2}_{\bf{\mathrm{m^{th}-order~transition}}}+ \nonumber \\	
	&~~~~~\underbrace{2\sum_{m'<m}\mathrm{Re}\left\{(U_{ji}^{m}(T))(U_{ji}^{m'}(T))^*\right\}}_{\mathrm{Interferences~between~different~transitions}}.
	\label{transprob}
\end{align}

The robustness criteria are formulated as follows: using the case of spectral amplitude uncertainty as an example, the amplitude pathways are first normalized with respect to the product of spectral amplitudes involved in the pathways as shown in (\ref{amppath}):
\begin{gather}
	c_{\vec{\alpha}} = \frac{U_{ji}(T,\vec{\alpha})}{\prod_{k}^K A_k^{\alpha_k}}.
	\label{amppathnorm}
\end{gather}
 Given the reasonable assumption that the amplitude modes $A_1,~\cdots,A_K$ are independent variables with uncorrelated distribution, the expected amplitude pathway $\mathrm{E}[U_{ji}(T,\vec{\alpha})]$ can be determined as follows:
\begin{align}
	\mathrm{E}[U_{ji}(T,\vec{\alpha})]&= c_{\vec{\alpha}}\prod_k^K\mathrm{E}[A_k^{\alpha_k}].
\end{align}
In turn, the $m$-th order contribution $\mathrm{E}[U_{ji}^m(T)]$ to the transition amplitude is:
\begin{align}
	\mathrm{E}[U_{ji}^m(T)]= \sum_{\vec{\alpha}\in \mathcal{M}}c_{\vec{\alpha}}\prod_{k}^K\mathrm{E}[A_k^{\alpha_k}],
\end{align}
where the sum term represents the addition of all amplitude pathways of order $m$. The expectation value of the total transition amplitude can be subsequently calculated as:
\begin{align}
	\mathrm{E}[U_{ji}(T)]&=\mathrm{E}\left[\sum_mU_{ji}^m(T)\right] \\ \nonumber
		&=\sum_{m}\sum_{\vec{\alpha}\in \mathcal{M}} \mathrm{E}[U_{ji}(T,\vec{\alpha})],
\end{align}
and the first moment of transition probability as:
\begin{align}
	\mathrm{E}&\left[P_{ji}(T)\right]=\mathrm{E}\left[\sum_m\left|U_{ji}^m\right|^2\right]+ \nonumber \\
	&\mathrm{E}\left[2\sum_{m'< m}\mathrm{Re}\left\{(U_{ji}^{m'})(U_{ji}^{m})^*\right\}\right],
	\label{momtransprob}		
\end{align}
where,
\begin{align}
	\mathrm{E}&\left[\sum_{m}\left|U^m_{ji}\right|^2\right]=\sum_{m}\sum_{\vec{\alpha}\in \mathcal{M}}\left|c_{\vec{\alpha}}\right|^2 \prod_{k}^K\mathrm{E}[A_k^{2\alpha_k}]  + \nonumber \\
	&~~~2\sum_m\sum_{\begin{subarray}{c}
	\vec{\alpha}' \in \mathcal{M} < \\
	\vec{\alpha}\in \mathcal{M}
	\end{subarray}}\mathrm{Re}\left\{c_{\vec{\alpha}'}c^*_{\vec{\alpha}}\prod_{k}^K \mathrm{E}[A_k^{\alpha_k'+\alpha_k}] \right\},
\end{align}
and
\begin{align}
	E&\left[2\sum_{m'< m}\mathrm{Re}\left\{(U_{ji}^{m})(U_{ji}^{m'})^*\right\}\right] = \nonumber \\
	&= 2\sum_{m'< m}\sum_{\begin{subarray}{c}
		\vec\alpha' \in \mathcal{M}, \\
		\vec\alpha \in \mathcal{M'}
		\end{subarray}} \mathrm{Re}\left\{c_{\vec{\alpha}'}c^*_{\vec{\alpha}} \prod_{k}^K\mathrm{E}[A_k^{\alpha_k+\alpha_k'}]\right\}.
	\label{mominterf}
\end{align}
The binary operator "$<$" applied to $\vec \alpha$ in the expressions above refers to any ordering of pathways, such as ${\vec \alpha' < \vec \alpha}$ if ${\alpha'_{k_{min}} < \alpha_{k_{min}}}$ where ${k_{min}\equiv \mathrm{min}~k~|~ \alpha_k' \neq \alpha_k}$. It is worthwhile to note that the calculation of the moment of transition probability involves interferences between pathways of the same and different order ( i.e. $c_{\vec\alpha}c^*_{\vec\alpha'}$ for ${(\vec\alpha, \vec\alpha') \in \mathcal{M}}$ and ${\vec\alpha \in \mathcal{M}}$ and ${\vec\alpha' \in \mathcal{M'}}$). The latter is specifically associated with the determination of the moment of interferences between transitions of different order. Both of these terms can be calculated for complete mechanistic analysis of quantum control robustness.
Additionally, the variance of the transition amplitude can be expressed as the following:
\begin{align}
	\mathrm{var}&\left(\mathrm{Re, Im}\left\{U_{ji}(T)\right\}\right)=\nonumber \\
	&=\mathrm{E}\left[\left(\sum_{\vec{\alpha}}\mathrm{Re,Im}\left\{c_{\vec{\alpha}}\right\}\left(\prod_{k}^K A_k^{\alpha_k} - \prod_{k}^K\mathrm{E}[A_k^{\alpha_k}]\right)\right)^2\right] \nonumber \\
	&=\hspace{0.1in}\sum_{\vec{\alpha}}\left(\mathrm{Re,Im}\left\{c_{\vec{\alpha}}\right\}\right)^2 \left(\prod_{k}^K\mathrm{E}[A_k^{2\alpha_k}]- \prod_{k}^K\mathrm{E}^2[A_k^{\alpha_k}]\right)+ \nonumber \\
	&~~~~~~2\sum_{\vec{\alpha}'< \vec{\alpha}}\mathrm{Re,Im}\left\{c_{\vec{\alpha}}\right\} \mathrm{Re,Im}\left\{c_{\vec{\alpha}'}\right\} \times \nonumber \\
	&~~~~~~\left(\prod_{k}^K\mathrm{E}[A_k^{\alpha_k'+\alpha_k}]-\prod_{k}^K\mathrm{E}[A_k^{\alpha_k'}]\prod_{k}^K\mathrm{E}[A_k^{\alpha_k}] \right).
	\label{vartransprob}
\end{align}
$\left[\mathrm{Re,Im}~U_{ji}\right]_{wc}$ can be obtained via equations (\ref{Jdistrib}) and (\ref{vartransprob}).
The expected transition probability is given by:
\begin{align}
	\mathrm{E}&\left[P_{ji}(T)\right] = \mathrm{E}\left[\left|\sum_{\vec{\alpha}} c_{\vec{\alpha}}\prod_{k}^K A_k^{\alpha_k}\right|^2\right] \nonumber \\
	&=\hspace{0.1in}\sum_{\vec{\alpha}} \left|c_{\vec{\alpha}}\right|^2 \prod_{k}^K \mathrm{E}[A_k^{2\alpha_k}] + \nonumber \\
	&2\sum_{\vec{\alpha}' < \vec{\alpha}} \mathrm{Re}\left\{c_{\vec{\alpha}'} c^*_{\vec{\alpha}}~\prod_{k}^K\mathrm{E}[A_k^{\alpha_k'+\alpha_k}]\right\}.
\end{align}
The expression for $\mathrm{var}~P_{ji}$ can be derived analogously.

A controller may choose to either arbitrarily specify the maximum Dyson order $M$ at first and check its accuracy based on the time order expansion of the \sd equation, or choose M based on the upper bounds on moment approximation errors. Continuing from (\ref{amppath}) and (\ref{amppathnorm}), the upper bound of the calculation is derived in the Appendix. The first moment of dipole pathway can be computed in an analogous fashion. The normalized dipole pathway is in turn given as:
\begin{gather}
       c_{\vec{\alpha}} = \frac{U_{ji}(T,\vec{\alpha})}{\prod_{p<q} \mu_{pq}^{\alpha_{pq}}},
\end{gather}
with $U_{ji}(T,\vec{\alpha})$ given in equation (\ref{dippath})
and the expressions for $\mathrm{E}[U_{ji}]$ and $\mathrm{var}(U_{ji})$ are identical to those for amplitude uncertainty,
with $\mathrm{E}[A_i^{\alpha_i}]$ replaced by $\mathrm{E}[\mu_{ji}^{\alpha_{ji}}]$. Here, the $\mu_{pq}$ correspond the elements of $\theta$ in (\ref{fisher}-\ref{genpost}).

Calculation of all moments of the control and system parameters can be computed once and used where they appear in the moment expressions (\ref{momtransprob}) and (\ref{vartransprob}). Given a particular distribution of a parameter, for e.g. amplitude mode $A_k$, the different moment terms $E[A_k^{\alpha_k}]$ can be computed. As an example, assuming $A_k, k\in[1,K]$ is Gaussian distributed, for a particular $k$ $E[A^{\alpha}]$ are calculated as follows:
\begin{align*}
	\mathrm{E}[A]&=\bar A, \\
	\mathrm{E}[A^2]&=\sigma^2 + \bar A^2,
\end{align*}
with the higher moment term $\mathrm{E}[A^\alpha]$ calculated recursively using the expression below starting with $\alpha=3$ as follows:
\begin{align*}
	\mathrm{E}[A^{\alpha}] &= \mathrm{E}[(A-\bar A)^{\alpha}]-\nonumber \\
	&~~~~\sum_{i=0}^n \left(
	\begin{array}{ccc}
		&\alpha&\\
		&&\\
		i&& (\alpha-i)
	\end{array}
		\right) \mathrm{E}[A^{i}](-\bar A)^{(n-i)},
		\label{momparam}
\end{align*}
\begin{align*}
	\mathrm{E}[(A-\bar A)^{\alpha}]&=
	\left\{
		\begin{array}{cl}
		0,&\mathrm{\alpha~odd}\\
		(\alpha-1)\sigma^{\alpha},&\mathrm{\alpha~even}
	\end{array}
	\right\}.
\end{align*}
This method can be extended to include other probability distributions, which can be expected to arise in different experimental conditions.

The approach for computing the quantum control robustness criteria described above assumes that the various order quantum pathways have been calculated and sorted in terms of the amplitude, phase and dipole parameters. In the case of dipole parameter uncertainty, the $E[\mu_{pq}]$ and the $\sigma^2(\mu_{pq})$ correspond to the parameter estimates and variance of parameter estimates (\ref{fisher}), in which $\Sigma$ is assumed to be diagonal. While these pathways could be evaluated by multiple integration of the Dyson terms, this can be computationally taxing especially when a large number of Dyson terms are involved in the dynamics. An efficient method to factorize the different contributions of the field's spectral parameters and system's dipole in the Dyson series is described in the next subsection. Convergence analysis of the aforementioned moment expressions will be presented in a separate work.

\subsection{Fourier encoding of control and system parameters}

The different quantum pathways defined by (\ref{amppath}) and (\ref{dippath}) can be efficiently computed using a commonly used method in signal processing, referred to as \emph{Fourier encoding/decoding}. In fact, due to the complexity of the explicit expressions (\ref{amppath}) and (\ref{dippath}) for the quantum pathways, it is convenient to define these pathways in terms of Fourier transforms. The technique was originally implemented to study the mechanism of controlled quantum dynamics \cite{mitr2003, mitr2008}.

In revealing amplitude pathways, a set of Fourier functions are implemented as \emph{amplitude encoding}:
\begin{gather}
	A_k \rightarrow A_k e^{\imath \gamma_k s}, \nonumber \\
	A^{\alpha_k}_k \rightarrow A^{\alpha_k}_k e^{\imath (\alpha_k\gamma_k) s},
\label{ampencode}
\end{gather}
where $\gamma_k$ is the \emph{modulating frequency} specific to the amplitude power $\alpha_k$ associated with a particular pathway ($\vec\alpha$). Using the modulation, the \sd equation can be propagated in the time variable $t$ and dummy variable $s$, for which the resulting \emph{encoded} transition amplitude $U_{ji}(T,s)$ is:
\begin{align}
	U_{ji}&(T,s)=\sum_{m}\left(\frac{\imath}{\hbar}\right)^m\sum_{\vec\alpha \in \mathcal{M}} e^{i(\sum_{k}^K \alpha_k\gamma_k)s} \prod_{k}^K A_k^{\alpha_k} ~\times \nonumber \\
	&\sum_{(k_1,\cdots,k_m)} \sum_{l_{m-1}}^N \mu_{jl_{m-1}} \int_0^T e^{i\omega_{jl_{m-1}}t_m}\cos(\omega_{k_m}t_m + \phi(\omega_{k_m})) \times \nonumber\\
	&\cdots\times \sum_{l_{1}}^N \mu_{l_11}\int_0^{t_2} e^{i\omega_{l_1i}t_1}\cos(\omega_{k_1}t_1 + \phi(\omega_{k_1})) ~ dt_1 \cdots dt_m.
\end{align}
The encoded total transition amplitude can be expressed in terms of amplitude pathways as:
\begin{align}
	U_{ji}(T,s)=\sum_{m}\sum_{\vec\alpha \in \mathcal{M}}U_{ji}(T,\vec{\alpha})e^{i(\sum_{k}^K\alpha_k\gamma_k)s}.
	\label{encoded}
\end{align}
Deconvolution of the total transition amplitude leads to
\begin{align}
	U_{ji}(T,\gamma) = \int_{-\infty}^{\infty} U_{ji}(T,s)e^{-i \gamma s}~ds.
	\label{decon}
\end{align}
This suggests that all amplitude pathways of different orders can be extracted through deconvolution of the encoded transition amplitude if all $\gamma$'s associated with each pathway is uniquely known, i.e. $U_{ji}(T,\gamma=\sum_{k=1}^K\alpha_k\gamma_k) \rightarrow U_{ji}(T, \vec \alpha)$. We can thus use (\ref{encoded}) along with (\ref{ampencode}) and (\ref{decon}) to concisely define amplitude pathways $\vec\alpha$ in (\ref{amppath}).

Similarly, \emph{dipole encoding} would reveal the contribution of the dipole moments in the transition amplitude. Here, each of the dipole matrix elements is encoded with a Fourier function:
\begin{gather}
	\mu_{pq} \rightarrow \mu_{pq}e^{\imath\gamma_{pq}s}, \nonumber \\
	\mu_{pq}^{\alpha_{pq}} \rightarrow \mu_{pq}^{\alpha_{pq}}e^{\imath(\alpha_{pq}\gamma_{pq})s},
\label{dipencode}
\end{gather}
with $\gamma_{qp}=\gamma_{pq}$.
The encoded and propagated unitary propagator consists of the different order dipole pathways with the encoded total transition amplitude:
\begin{align}
	U_{ji}&(T,s)=\sum_{m} \sum_{\vec\alpha \in \mathcal{M}}U_{ji}(T,\vec{\alpha})e^{i\left(\sum_{p<q}\alpha_{pq}\gamma_{pq}\right) s}.
	\label{dipdecon}
\end{align}
Deconvolution of the total transition amplitude leads to the decoded dipole pathway, i.e. $U_{ji}(T,\gamma=\sum_{p<q} \alpha_{pq}\gamma_{pq}) \rightarrow U_{ji}(T,\vec{\alpha})$. We can similarly use (\ref{dipdecon}) along with (\ref{dipencode}) and (\ref{decon})  to define dipole pathways in (\ref{dippath}). Now that the contribution of the control and system parameters to the different orders of the Dyson terms have been delineated, this information together with moments of parameters, can be used to explicitly calculate the effect of manipulated input 
or system parameter uncertainties on the quantum interferences and transition probability. The details of the numerical implementation of the method is discussed in the next section.

\subsection{Worst-case robustness analysis}
\label{Section43}
As noted above, worst-case robustness analysis can also be carried out based on constrained maximization of the distance between the nominal and worst-case values of the performance measure \cite{kosu2013}. These approaches are based on leading order Taylor expansions.
For example, in a first-order formulation, the problem can be expressed as
\begin{align}
	 \underset{\delta\theta \in \Theta}{\mathrm{max}}~|\delta J|^2  \approx \delta \theta^T (\nabla_\theta J)^T \nabla_\theta J \delta \theta,
	 \label{deviate}
\end{align}
where $\Theta$ was defined in (\ref{distrib}) and $\nabla_\theta J$ in (\ref{gradJ}) (assuming $J=P_{ji}$). If we let $x=\chi_K^{-1}(c) Q^{-1} \delta \theta$, where $Q^TQ=\Sigma$, then under this change of variables the constrained maximization problem (\ref{deviate}) is mapped:
\begin{align}
	\underset{\delta \theta \in \Theta}{\mathrm{max}}~ |\delta J|^2 \rightarrow \underset{x^Tx \leq 1}{\mathrm{max}} \chi_{K}^2(c) x^T Q^T (\nabla_\theta J)^T (\nabla_\theta J) Q x.
\end{align}
This problem has the form of a Rayleigh quotient \cite{golu1996}, which has an analytical solution for $J_{wc}$ and $\theta_{wc}=\hat \theta + \delta \theta_{wc}$, with $\delta \theta_{wc} = \mathrm{arg~max}~|\delta J|^2$ written in terms of a singular value decomposition with appropriately chosen sign. However, since the formulation is first order, it is subject to the same issues of accuracy noted above. Future work will compare the accuracy of various approaches to estimation of $J_{wc}$ for quantum control systems.
%
\section{Numerical implementation}
\label{Section5}
\subsection{Fourier Encoding}

The key to a successful encoding (and, therefore, decoding) of quantum pathways is to ensure that the encoding frequency $\gamma$ for each pathway is unique. Hence, the choice of $\gamma_k$ directly depends on the pathway definition as given in (\ref{amppath}) and (\ref{dippath}) for amplitude and dipole pathways, respectively. For amplitude encoding, assuming that the significant number of Dyson terms is $M$, the encoding frequency corresponding to each amplitude mode must be separated by at least M terms. If $A_1$ is encoded with frequency $\gamma_1$, $A_2$ must be encoded with $\gamma_2=(M+1)\gamma_1$, and $A_k$ with $\gamma_k=(M+1)^{k-1}\gamma_1$. This is to ensure that each amplitude mode $A_k$ with power up to $M$ would not have overlapping encoding frequency with the rest of the amplitude modes $A_{k'},~k\neq k'~\epsilon~[1,K]$. As described in the previous section, the same set of encoding frequencies can be employed for the case of dipole pathways. Again, if the quantum dynamics is significant up to a Dyson order $M$, each transition between the intermediate states of $|i\rangle$ and $|j\rangle$ can be repeated at the most $M$ times, such that the dipole moment $\mu_{ij}$ would have a maximum power of $M$. Using this assumption, each $\mu_{ij},~i\neq j~\epsilon~[1,N]$ must be separated by M terms. For instance, if $\mu_{1j}=\mu_{j1}$ is encoded with $\gamma_{1j}=\gamma_{j1}$, then $\mu_{2j}=\mu_{j2}$ is encoded with $\gamma_{2j}=\gamma_{j2}=(M+1)\gamma_{1j}$ and $\mu_{nj}=\mu_{jn}$ with $\gamma_{nj}=\gamma_{jn}=(M+1)^{n-1}\gamma_{1j}$ for $n\neq j~\epsilon~[1,N]$. This type of encoding assumes that there is connectivity in all of the states within the quantum system (i.e. $\mu_{ij} \neq 0$ for all $i\neq j~\epsilon~[1,N]$). For a sparse dipole matrix, it may be more computationally efficient to start with an evenly spaced encoding frequency and subsequently ensure that none of them overlap during the decoding process.

\subsection{Fourier Decoding}
The decoding procedure begins with deconvolution of encoded transition amplitude via Fourier transform. Each deconvoluted term is then assigned to the appropriate pathway based on their respective sum of encoding frequencies. As discussed in the previous subsection, the encoding frequencies are initially chosen so that the $\gamma$'s for each pathway belonging to each order are implicitly known. This means that any pathways associated with $A_1^{\alpha_1},\cdots,A_K^{\alpha_K}$ are associated with $\gamma =\alpha_1\gamma_1+ \cdots + \alpha_K(M+1)^{K-1}\gamma_1$. Using this information, each sum of encoding frequency $\gamma$ is \emph{factorized} with respect to $\gamma_1$ to reveal all amplitude pathways of all orders, i.e. $U_{ji}(T,{\gamma=\sum_{k=1}^{K} \alpha_k\gamma_k}) \rightarrow U_{ji}^m(T,\vec{\alpha})$. The result is a set of amplitude pathways of up to a maximum order $M$. An analogous approach can be used for dipole pathways.

\section{Results: Example}
\label{Section6}
This section demonstrates the application of methods and procedures described in Section \ref{Section3}, \ref{Section4} and \ref{Section5} on an artificial quantum system.

The majority of quantum robust control studies - especially in the context of Hamiltonian uncertainty - have considered robustness of controlled quantum gate fidelity.
For example, gate control systems including qubit arrays with Heisenberg couplings \cite{heul2010}, atomic lattices \cite{khan2012},  as well as other coupled qubit systems
\cite{wang2012}  have been studied either from the perspective of the robustness of nominally optimal control fields (i.e., fields that were optimized in the absence of uncertainty), or the perspective of optimization in the presence of uncertainty.

The theory and methodologies developed in the present work are applicable to both control of population transfer in molecular systems, which is typically achieved using shaped femtosecond laser pulses \cite{levi2001}, and control of quantum gates.
Robustness analysis and robust control methods are especially important in laser control because there is currently no way to use real-time feedback methods to regulate the controlled dynamics. Thus far, successful laser control of molecular dynamics has been achieved almost exclusively through experimental learning loops that are not based on first-principles quantum mechanical models of the molecular systems. Model-based control techniques have not yet been successfully applied. Hence we emphasize laser controlled population transfer problems in our analysis and examples. Potential applications of our methods include model-based dynamic control of chemical reactions. In these applications, the robustness of quantum interferences between transition pathways is of particular importance.

The Hamiltonian parameters of the example system studied in the present work are chosen as follows:
\begin{align}
	H_{0}=\begin{pmatrix} 0 & 0 & 0 & 0 \\0 & 1 & 0 & 0 \\0 & 0 & 1.5 & 0 \\0 & 0 & 0 & 2 \end{pmatrix},~\mu=\begin{pmatrix} 0 & 2  & 1 & 0 \\2  & 0 & 0 & 2 \\1 & 0 & 0 & 0 \\ 0 & 2 & 0 & 0 \end{pmatrix}.
\end{align}
The system evolves according to (\ref{SDunit}), with the time-varying electric field $\varepsilon(t)$ parametrized as a linear combination of cosine waveforms. The manipulated field parameters are the spectral frequency, amplitude and phase, and the control objective is the maximization of the transition probability between the initial state $|1\rangle$ and the target state $|4\rangle$, i.e. i.e. $P_{41}(T)$.

Several types of optimization algorithms have been applied to identify control strategies that maximize the fidelity of controlled quantum dynamics in the presence of system or input field uncertainty, both for quantum gates \cite{stef2007,kosu2013,cabr2011} and control of observables \cite{beum1992, bart2005}.  For example, \cite{stef2007} considered microwave control of quantum gates in the presence of both pulse amplitude and frequency detuning, and proposed techniques for combating both simultaneously through a numerical optimization scheme. \cite{kosu2013} applied nonlinear programming algorithms to the design of robust quantum gate controls in the presence of system parameter uncertainty. These algorithms, commonly applied in engineering robust control, are well-suited to the solution of robust optimal control problems in the presence of constraints.
\cite{beum1992, bart2005} presented algorithms for identifying robust control solutions in the context of laser control of molecular dynamics.

Here, RCGA\footnote{RCGA is a stochastic optimization algorithm whose principle of optimality and convergence is based on survival of the fittest and principles of genetics. For more information regarding the procedure of the algorithm, the reader is encouraged to refer to \cite{gold1989, deb1995}} is employed to obtain the combinations of field parameters which maximize the objective. The decision variables of the optimization is formulated as $\vec{\mathrm{x}} \equiv [\omega_1, \cdots, \omega_K, \phi_1, \cdots, \phi_K]$, where the number of modes $K$ has been pre-determined to be 3. The field duration $T$ and the amplitude modes $A_k$ have also been pre-determined to be 10 and 0.1, respectively, based on value pre-screenings to ensure control optimality (data not shown). Table \ref{RCGAparam} summarizes the RCGA algorithmic parameters used to obtain the control solutions. The acquired control parameters are listed in Table \ref{ControlParam} and are analyzed for robustness below.
\begin{table}
\caption{\label{RCGAparam}RCGA algorithmic parameters used for obtaining quantum control fields listed in Table \ref{ControlParam}.}
\begin{ruledtabular}
\begin{tabular}{cc}
 	Operator	&Parameter \\ \hline
 	Initial Population	&Size=300 \\
 	Reproductive Population	&Size=30 \\
 	Crossover	&SBX, probability=0.2\\
 	Mutation	&Gaussian, probability=0.01 \\
	Selection	&Tournament, size=2 \\
\end{tabular}
\end{ruledtabular}
\end{table}
\begin{table}
\caption{\label{ControlParam}Quantum control field parameters obtained via RCGA optimization. The control field duration is fixed at $T=10$ and the amplitude modes $A_k$ at 0.1. The number of modes used in the optimization is 3.}
\begin{ruledtabular}
\begin{tabular}{ccc}
 	Index	&Frequency modes($\omega_k$)	&Phase modes($\phi_k$)	 \\ \hline
 	$\varepsilon_1$	&[1.0311, 2.4347, 1.0540]	&[3.6380, 3.3807, 3.4839]  \\
 	$\varepsilon_2$	&[1.7671, 1.0048, 1.0019]	&[4.7794, 4.2516, 4.2667]  \\
 	$\varepsilon_3$	&[1.0076, 1.0105, 1.7279]	&[1.1894, 1.1694, 1.8371]  \\
 	$\varepsilon_4$	&[1.0004, 1.0996, 1.0411]	&[2.3030, 3.3381, 3.5704]  \\
	$\varepsilon_5$	&[1.0067, 1.8850, 1.0426] &[0.6550, 0.6656, 0.4101]  \\
	$\varepsilon_6$	&[3.7307, 1.0442, 1.0209]	&[0.0449, 0.4724, 0.6493]  \\
	$\varepsilon_7$	&[3.0631, 1.0239, 1.0512]	&[0.2068, 0.5943, 0.4091]  \\
	$\varepsilon_8$	&[1.0009, 1.0112, 1.8064] &[0.8815, 0.8174, 1.3002]  \\
\end{tabular}
\end{ruledtabular}
\end{table}
\begin{figure}
	\centering
	\includegraphics[width=8.2cm]{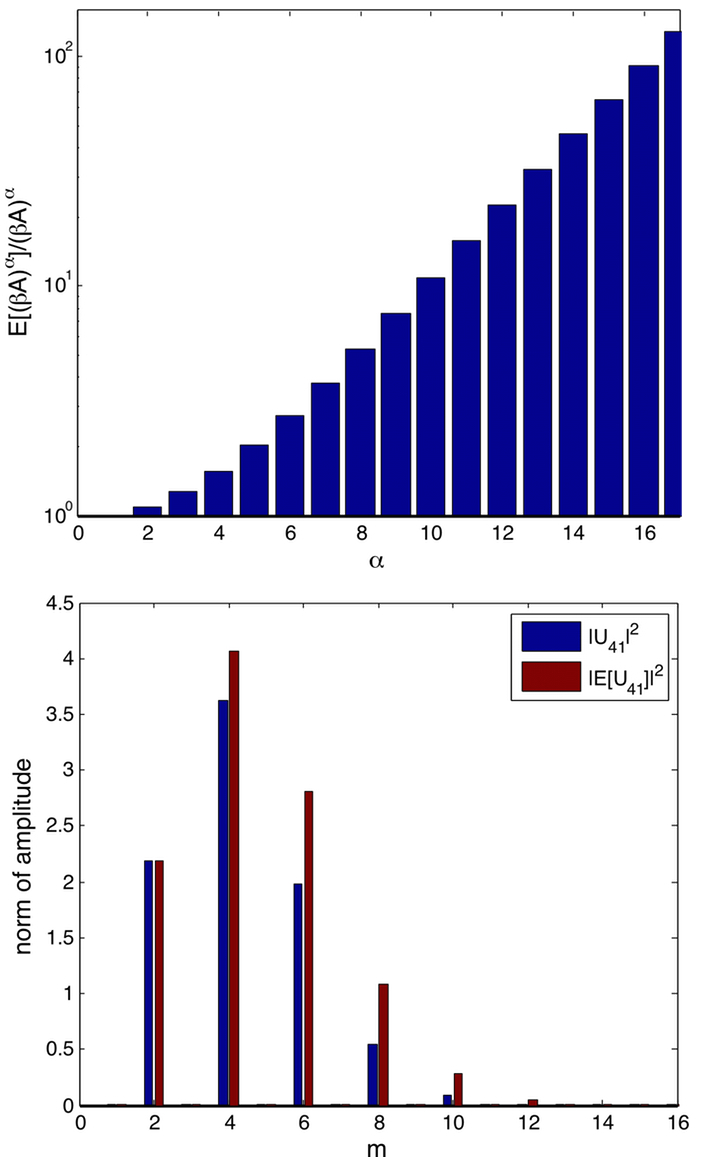}
	\caption{(Top) Calculation of the first moment of $A^\alpha$($E[A^\alpha]$). The log plot shows that for an amplitude mode with a Gaussian distribution the moment of amplitude with increasing power becomes exponentially larger relative to its nominal value. (Bottom) The bar plot shows the Dyson terms involved in $\e_1$ at nominal and expected case ($\sigma(A_k)$=0.3). It suggests that terms of higher orders, which may not be negligible under nominal condition will become significant in a noisy environment.}
	\label{normamplitudexpectation}
\end{figure}	

In this example, the contribution of Gaussian uncertainty
in the spectral amplitude is considered. Nevertheless, as discussed in previous sections, an analogous analysis can be readily performed in the case of dipole parameter distribution. As described in Section \ref{Section4}, the robustness analysis reveals how distribution in the parameters due to uncertainty
affects the pathway interference and transition probability. Traditionally, as in classical control, the robustness criteria have been defined as the first and second moment of the control objective. While these criteria are directly applicable in the quantum case, moments of pathway interferences provide additional insights into the mechanism of quantum control robustness. The theoretical and numerical implementation of the robustness analysis are performed using the procedure described in section \ref{Section4} and \ref{Section5}, respectively. In the first step, modulation of the \sd equation using Fourier functions are performed to reveal the amplitude pathways. The Fourier encoding parameters corresponding to the three amplitude modes obtained in the optimization are $\gamma_1=1$, $\gamma_2=22$ and and $\gamma_3=485$, respectively. Post-encoding, the encoded unitary propagator is deconvoluted and the resulting decoded matrix is identified as a particular pathway according to its encoding frequency. Table \ref{PathNomExp} lists the significant amplitude pathways of the first two order sorted in terms of their order and decoded frequency.
\begin{figure}
	\includegraphics[width=9cm]{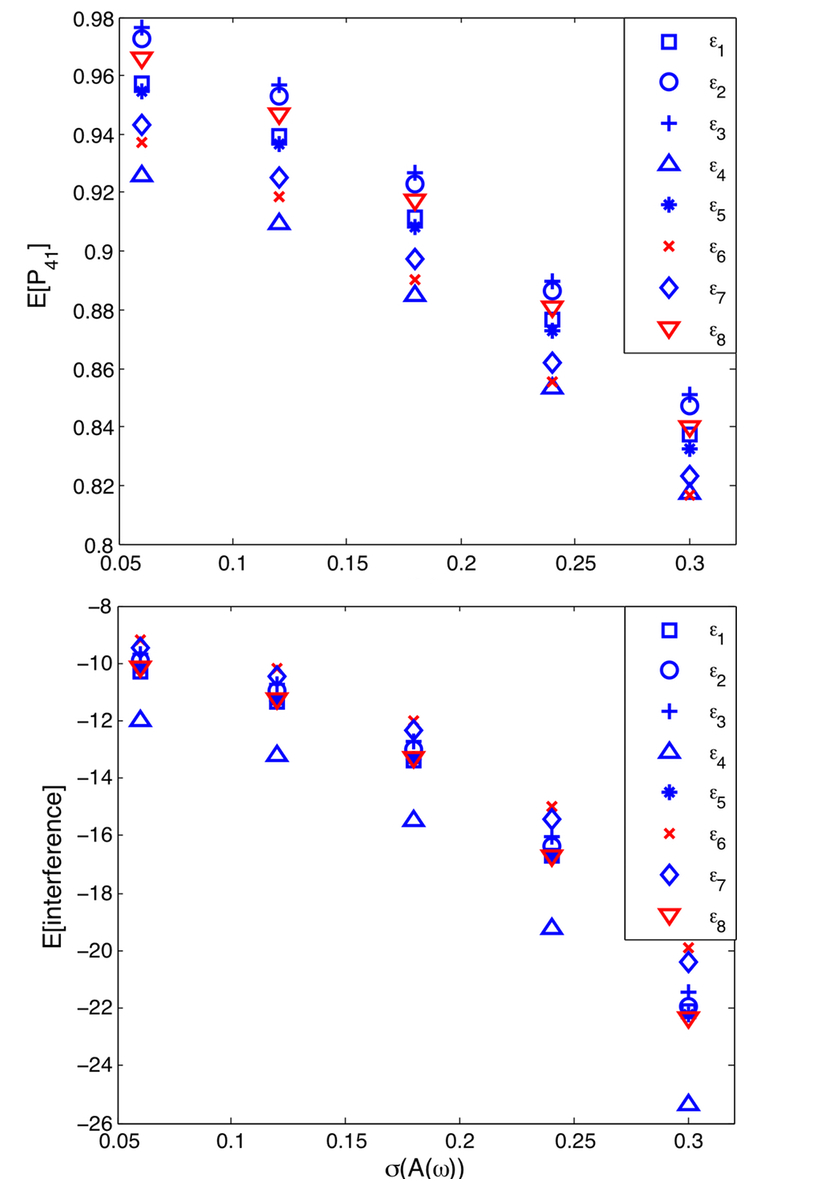}
	\caption{Plot of expected population transfer (top) and interference (bottom). The data suggests that control field implementation uncertainty increases destructive interference and in turn reduces the moment of transition probability.}
	\label{interfnoi}
\end{figure}
\begin{figure}
	\includegraphics[width=9cm]{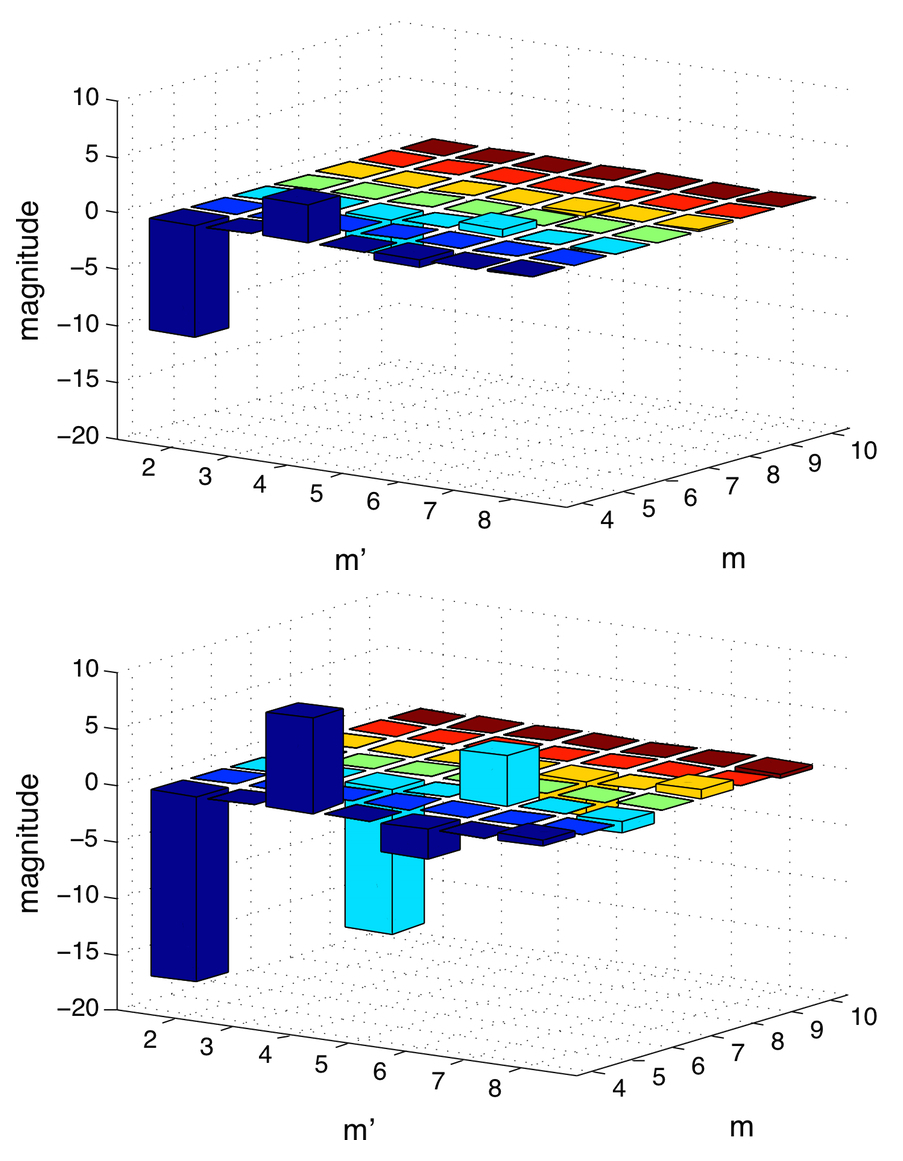}
	\caption{Bar plot of interferences between different pathways involved in control field dynamics of $\varepsilon_1$ from Table \ref{ControlParam} in nominal (top) and expected case (bottom) ($\sigma(A_k)=0.3$). As seen from the plot, Gaussian uncertainty
increases the destructive interferences between transitions.}
	\label{interf_nomvsexp}
\end{figure}
\begin{figure}
	\centering
	\includegraphics[width=9cm]{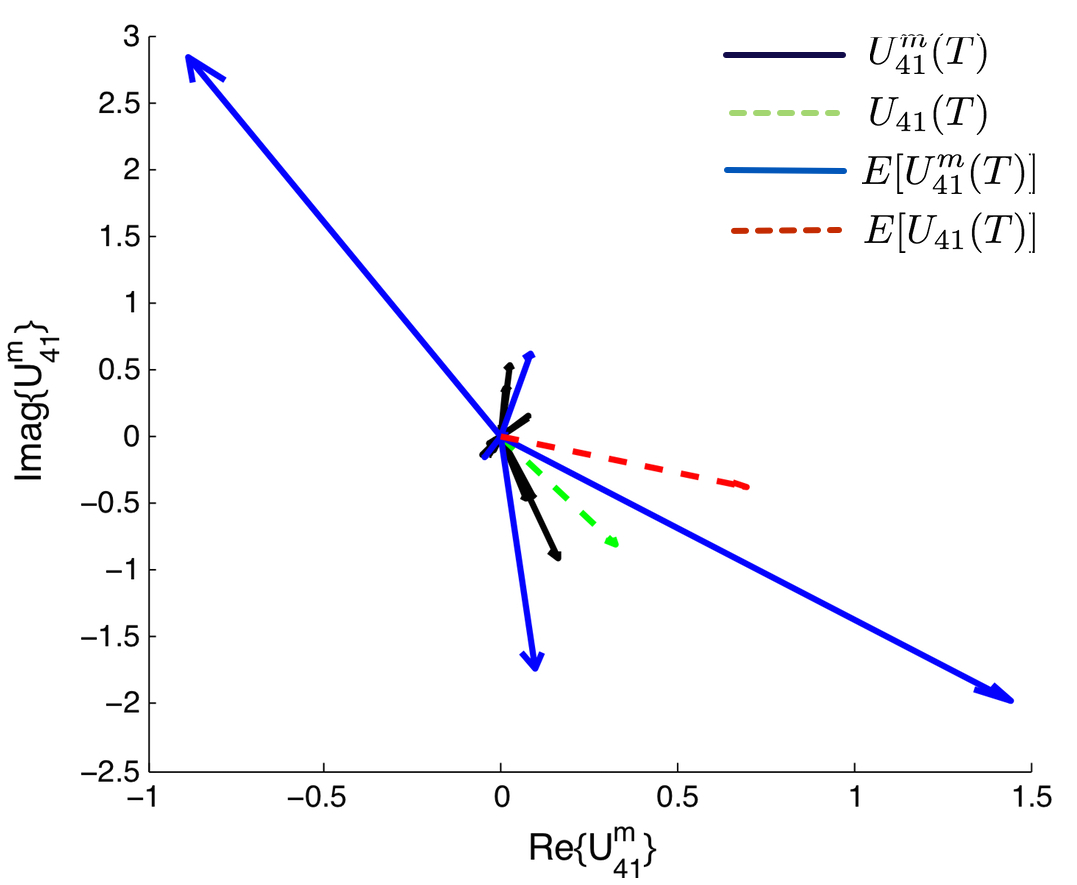}
	\caption{Plot of nominal and expected transition amplitude ($\sigma(A_k)=0.3$) associated with $\e_1$ from Table \ref{ControlParam} decomposed in terms of each Dyson term.}
	\label{amplitudexpectation}
\end{figure}
\begin{figure*}
	\includegraphics[width=18cm]{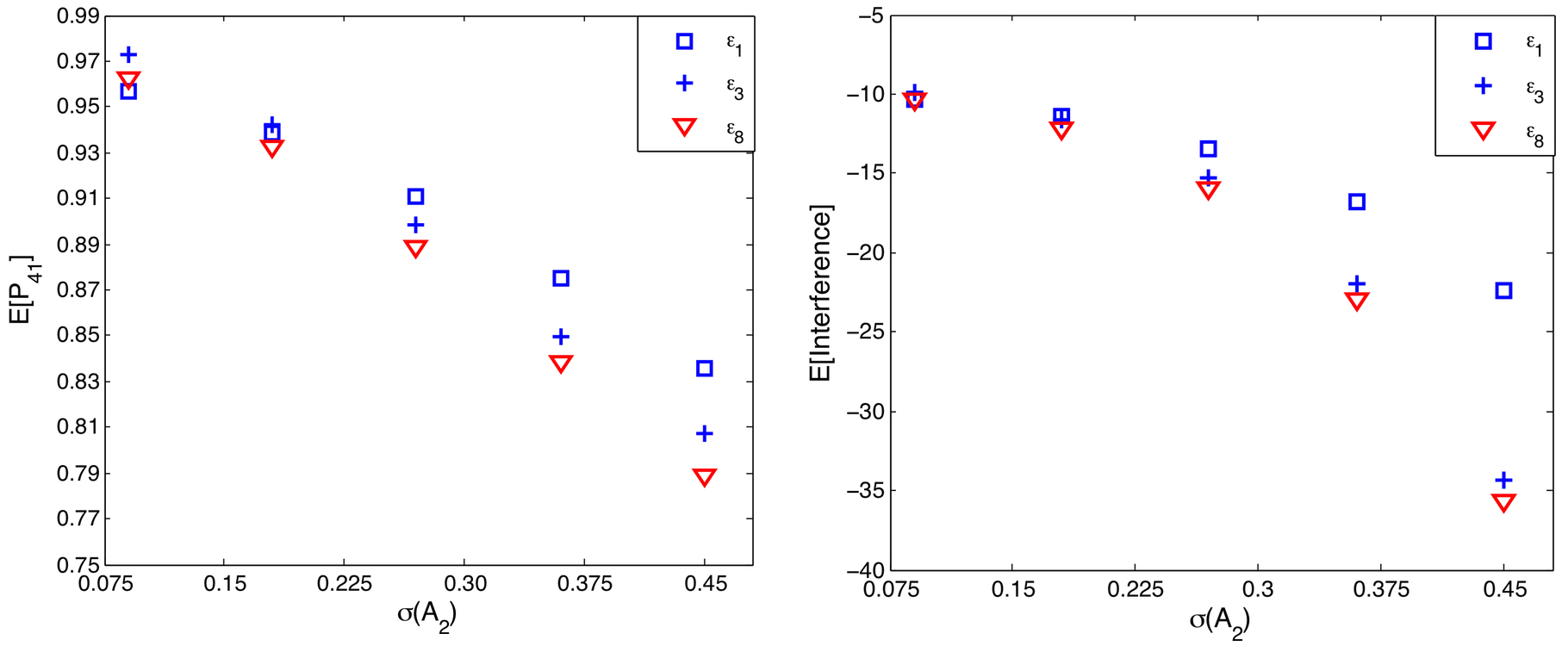}
	\caption{Plot of expected population transfer (left) and interference (right) for
uncertainty distributions that differ across amplitude modes. The $x$- axis shows increasing standard deviation of the second amplitude mode, while the rest of the amplitude modes are fixed at 0.3. The data suggests that certain pathways are more resistant to uncertainty leading to higher expected transition probability and reduced destructive interferences.}
	\label{robustfieldinterfnoi}
\end{figure*}
\begin{figure*}
	\includegraphics[width=18cm]{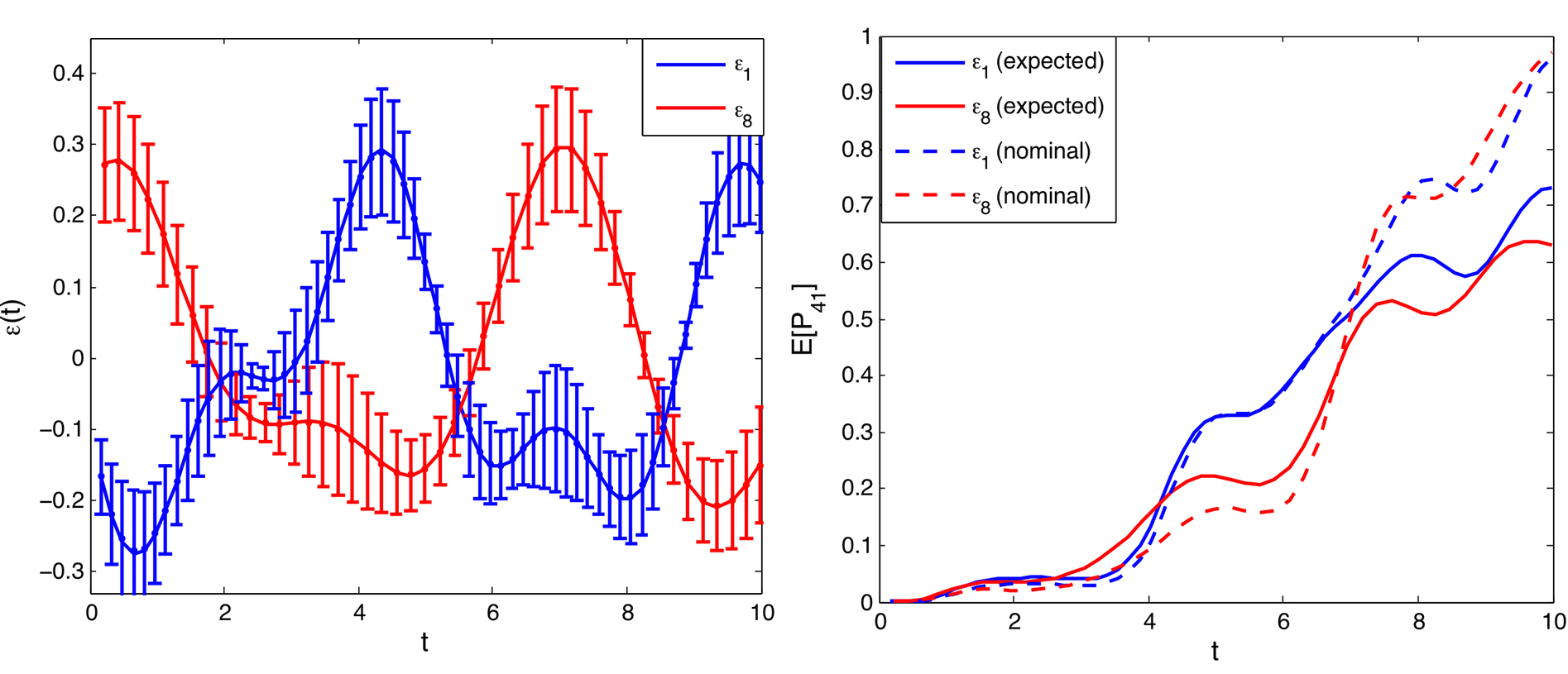}
	\caption{Plot of temporal field of $\varepsilon_1$ and $\varepsilon_8$ as defined in Table \ref{ControlParam} (left) and their corresponding population transfer trajectory (right) in the nominal and expected case ($\sigma(A_1)=\sigma(A_2)=0.45$ and $\sigma(A_2)=0.65$).}
	\label{robustfieldnoi}
\end{figure*}
\begin{table*}
\caption{\label{PathNomExp} The table shows the amplitude pathways of the first two significant order ($m=[2,4]$) associated with $\varepsilon_1$ from Table \ref{ControlParam} with their corresponding encoding frequency $\gamma$ and contribution to the transition amplitude in the nominal and expected case with variance ($\sigma(A_k)=0.3$).}
\begin{ruledtabular}
\begin{tabular}{p{2.2cm}p{2.2cm}cp{2.3cm}p{2.3cm}p{2.3cm}}
 	Pathway($\vec{\alpha}=[\alpha_1,\alpha_2,\alpha_3]$)	&Encoding frequency($\gamma$)	&Amplitude	&$U_{41}^m(T)$	&$E[U_{41}^m(T)]$	&$\mathrm{Var}~(\mathrm{Re}\left\{U^m_{41}\right\})+\mathrm{Var}~(\mathrm{Im}\left\{U^m_{41}\right\})$\\\hline
 	$[0, 0, 2]$		&2 		&0.1816- 0.2149i		&	&	&\\
	$[0, 1, 1]$		&23		&0.7366 - 0.8381i		&	&	&\\
	$[0, 2, 0]$		&44		&0.7462 - 0.8164i		&0.2346 - 2.3926i	&0.2498 - 2.4861i 	&0.0190+0.9171i\\
	$[1, 0, 1]$		&485		&-0.04693 - 0.07756i	&	&	&\\ 	
	$[1, 1, 0]$		&506		&-0.09012 - 0.1555i		&	&	&\\ \hline
 	$[0, 0, 4]$		&4		&-0.01389 + 0.02558i	&	&	&\\	
	$[0, 1, 3]$ 	&25		&-0.1101 + 0.2025i		&	&	&\\
	$[0, 2, 2]$		&46		&-0.3281 + 0.6014i		&	&	&\\
	$[0, 3, 1]$		&67		& -0.4356 + 0.7946i		&	&	&\\
	$[0, 4, 0]$		&88		&-0.21739 + 0.3941i		&0.3342 + 2.0936i	&0.3902 + 2.6029i	&0.1321+0.2991i\\
	$[1, 0, 3]$		&487		&0.01617 + 0.01658i		&	&	&\\
	$[1, 1, 2]$		&508		&0.09452 + 0.09914i		&	&	&\\
	$[1, 2, 1]$		&529		&0.1841 + 0.1979i		&	&	&\\
	$[1, 3, 0]$		&550		&0.1194 + 0.1318i		&	&	&\\
\end{tabular}
\end{ruledtabular}
\end{table*}
\begin{figure*}
	\includegraphics[width=18cm]{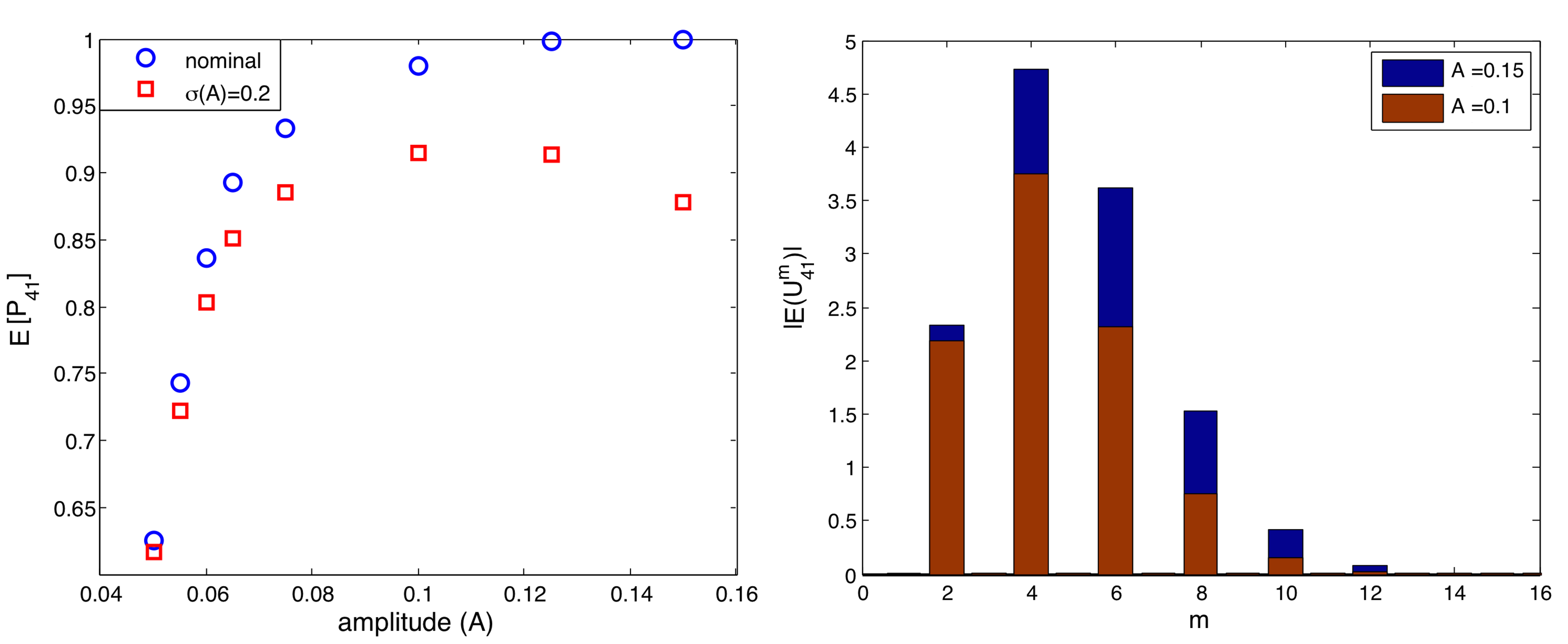}
	\caption{(Left) Plot of the expectation value of transition probability vs. increasing amplitude strength. The figure shows that, under nominal condition better control is achieved as stronger fields are optimized (\emph{circle}). In the presence of amplitude uncertainty
($\sigma=0.2$), however, there is a particular field strength corresponding to a robust point which is not necessarily optimal under nominal condition (\emph{square}). (Right) The bar plot shows the significant number of Dyson terms involved in the dynamics of two control fields with different field strength. It indicates that the more robust field involves fewer significant Dyson terms and, hence, amplitude pathways relative to the non-robust field.}
	\label{complexityvsrobustness}
\end{figure*}

The next step of the analysis is the calculation of expected amplitude modes assuming a Gaussian parameter distribution. As shown in Figure \ref{normamplitudexpectation} (top), the ratio of the expected to nominal amplitude increases exponentially. This implies that higher order pathways which may be negligible under nominal condition would become significant in a noisy environment. Figure \ref{normamplitudexpectation} (bottom) shows an example in the case of control field $\e_1$ under nominal and noisy condition ($\sigma(A_k)=0.3$). In addition, calculation of the first moment of amplitude pathway shows how pathways of different orders change in magnitude and direction as the control parameter is distributed (Table \ref{PathNomExp}). The interferences between different pathways are calculated according to (\ref{mominterf}) for the nominal and noisy case. The results of the interference calculations are shown in Figure \ref{interfnoi} and they suggest that implementation inaccuracies 
destroy the constructive interference or amplify the destructive interference. This effect on destructive interference increases as parameter distribution is increased (Figure \ref{interf_nomvsexp} and Figure \ref{amplitudexpectation}) and reduces the control field's fidelity proportionally. The calculation for the second moment of transition amplitude is also performed and can be compared with the simulated values (Table \ref{variance}). The trend is analogous to that of the first moment in that the magnitude of the variance increases as the variance of the amplitude modes increases. These values can be further used for the calculation of worst-case scenario $J_{wc}$ as discussed in Section \ref{Section41} in (\ref{Jdistrib}).
\begin{table*}
\caption{\label{variance} The table lists the expected transition probability and variance of transition amplitude associated with the control field $\varepsilon_1$ defined in Table \ref{ControlParam} as correlated with increasing variance of the field amplitude disturbance distribution. The computed values are compared to its simulated counterpart (out of 800 samples).}
\begin{ruledtabular}
\begin{tabular}{p{1cm}p{1.5cm}p{2cm}p{3.5cm}p{3.5cm}}
 	$\sigma(A(\omega_k))$	&E[$P_{41}$]	&mean($P_{41}$) (by sampling)		&$\mathrm{Var}~(\mathrm{Re}\left\{U_{41}\right\})+\mathrm{Var}~(\mathrm{Im}\left\{U_{41}\right\})$ 	&$\mathrm{Var}~(\mathrm{Re}\left\{U_{41}\right\})+\mathrm{Var}~(\mathrm{Im}\left\{U_{41}\right\})$ (by sampling) \\ \hline
 	0.06	&0.9571	&0.9550	&8.295e-4+6.968e-005i		&9.323e-4+9.243e-005i	\\	
 	0.12	&0.9392	&0.9345	&0.003163+0.0005661i		&0.003388+0.0006441i	\\
	0.18	&0.9115	&0.9089	&0.006558+0.002185i		&0.007091+0.002265i	 \\
	0.24	&0.8766	&0.8688	&0.01038+0.005571i			&0.01165+0.008212i	\\
 	0.30	&0.8374	&0.8372	&0.01397+0.01072i			&0.01587+0.01361i		\\
\end{tabular}
\end{ruledtabular}
\end{table*}

In the laboratory, there are cases where the pdf is not identical across different input or system parameters. In this case, different pathways are affected by input or system parameter uncertainty to different extents. Under this condition, the amplitude robustness analysis showed that there exists a set of pathways which are less affected by implementation inaccuracies and thus, more robust. For this analysis, the standard deviations of the second of the three amplitude modes of the control fields listed in Table \ref{ControlParam} are varied while the rest are fixed at 0.3. The robustness analysis shows that some combination of amplitude modes and therefore, pathways are more resistant to implementation uncertainty, which in turn minimizes the effect of parameter distribution on destructive interference ($\e_8$) relative to its non-robust counterpart ($\e_1$ and $\e_7$) (Figure \ref{robustfieldinterfnoi}). As an illustration, Figure \ref{robustfieldnoi} shows the plot of relatively robust and non-robust control field ($\e_8$ and $\e_1$, respectively) and their corresponding control trajectories under nominal and noisy condition.

Moreover, given the multiplicities of quantum control solutions, we investigated how stronger fields which utilize more pathways are affected by
implementation uncertainties relative to their weaker counterparts. As seen in (\ref{DysSer}), fields with high amplitude and longer duration involve more Dyson terms and, as a result, more pathways. This subsequently poses more entry points for control
or system parameter uncertainty to affect the control's optimal state trajectory. To demonstrate this property, we perform optimization of control fields with variable time duration in order to analyze the optimality and robustness of the control as a function of field strength and number of Dyson terms involved in the dynamics. The same algorithmic parameters and decision variables as the ones used to obtain control fields listed in Table \ref{ControlParam} applies in this case, but with the amplitude strength across all three modes varied in a range between 0.05 to 0.15. Ten optimization runs were performed for each amplitude case and the best one is reported in Figure \ref{complexityvsrobustness} (left). As shown in the plot, fields with higher amplitude are better at maximizing transition probability under nominal conditions but perform worse when
implementation uncertainties are present. This observation is consistent with the interpretation of the robustness analysis, which is that while stronger fields utilize more quantum pathways and therefore results in greater control, these fields may become more susceptible to implementation errors manifests in more pathways (Figure \ref{complexityvsrobustness} (right)).
This observation also suggests that there is a trade-off between the optimality of a control field and its robustness.

These results demonstrate that a) nominally optimal controls are generally not the most robust and do not provide the highest expectation values of controlled observables; b) the mechanistic origin of the reduced robustness of nominally optimal controls lies in their use of higher order pathways and associated quantum interferences, which are sensitive to uncertainty. In this regard, \cite{negr2011} reported methods for the design of simple, easy-to-implement control pulses that are close-to-optimal but not necessarily optimal. The above analysis shows why easier-to-implement control pulses may be more robust, and provides theoretical foundations for the identification of controls that employ such robust population transfer mechanisms.

\section{Summary and Prospective}
\label{Section7}
In this paper, theoretical foundations for the robustness analysis of coherent quantum control systems have been presented.
This theory enables prediction of moments of observables in any bilinear coherent control system under Hamiltonian parameter and input field uncertainty, without the use of leading order approximations.
Due to the bilinear nature of the interaction between a quantum system and an external field, the dynamics of controlled quantum system can be described using Dyson expansion. The resulting Dyson terms can in turn be interpreted as a combination and interference of quantum pathways, appropriately defined for the purpose of robustness analysis, whose outcome is a transition amplitude and probability between an initial and final state. These pathways are an explicit function of the control and system parameters such that the effect of control field implementation errors
and system parameter uncertainty on the state-to-state transitions can be calculated using the expressions and associated computational methodologies derived herein. The robustness criteria of controlled quantum dynamics include the moment of quantum control objectives, such as the transition amplitude and probability. Moreover, since quantum pathways interfere with one another in order to produce the observed dynamics, the moment of interference is an essential robustness criteria in the understanding of quantum control robustness.

The robustness analysis method described herein can be implemented in robust control algorithms in a couple of ways. First, robustness analysis for time-independent Hamiltonian uncertainty can be used to compute model-based robust control solutions in an open-loop setting given Hamiltonian parameter estimates. This can in turn be used in conjunction with deterministic robust control algorithms to achieve robust solutions based on either distributional or worst-case criteria, specifically, taking into account quantum pathway interferences and maximizing the performance measure by minimizing the destructive interference. Future work may also compare the mechanisms by which robustness is achieved using the present methodology with those obtained under leading order approximations, for problems with more general types of correlation in equations (\ref{fisher}) and (\ref{freqcorr}). Second, the asymptotic nature of robustness analysis can be used to help determine the number of observations required to obtain accurate estimates of control robustness using experimental sampling of noisy fields in learning control algorithms. These robust open-loop and learning control methods could ultimately be combined in model-based quantum adaptive feedback control. Finally, these asymptotic methods are also applicable to robustness analysis of other bilinear systems and may prove useful in robust control of such systems. However, it is important to emphasize that the role of interferences (and robustness thereof) in producing the observed robustness of quantum dynamics is unique to quantum control.

\section{Acknowledgements}
R. C. developed the theory; A. K. carried out simulations. R. C. and A. K. wrote the paper. Financial support was provided by the Department of Chemical Engineering, Carnegie Mellon University.


\end{document}